\documentclass[journal]{IEEEtran}
\makeatletter
\def\endthebibliography{%
	\def\@noitemerr{\@latex@warning{Empty `thebibliography' environment}}%
	\endlist
}
\makeatother

\usepackage[noadjust]{cite}

 \usepackage[pdftex]{graphicx}

\ifCLASSINFOpdf
\else
\fi

\usepackage{algorithmic}
\usepackage{algorithm}

\usepackage{array}
\newcolumntype{K}[1]{>{\raggedright\arraybackslash}m{#1}} 
\newcolumntype{X}[1]{>{\centering\arraybackslash}m{#1}} 
\newcolumntype{E}[1]{>{\raggedleft\arraybackslash}m{#1}} 

\usepackage{amsmath} 

\usepackage[table]{xcolor} 

\usepackage{textcomp} 

\usepackage[utf8]{inputenc}

\begin{document}
%
\title{Coverage Path Planning for Thermal Interface Materials}
%
%
%

\author{Simon~Baeuerle, Andreas~Steimer* and Ralf~Mikut*
\thanks{The work of R. Mikut was supported by the Helmholtz Association’s Initiative and Networking Fund through Helmholtz AI and the program Energy System Design. \textit{(* A. Steimer and R. Mikut contributed equally to this work.) (Corresponding author: S. Baeuerle.)}}
\thanks{S. Baeuerle and R. Mikut are with the Institute for Automation and Applied Informatics (IAI), Karlsruhe	Institute of Technology (KIT), D-76344 Eggenstein-Leopoldshafen, Germany. (e-mail: simon.baeuerle@kit.edu)}
\thanks{S. Baeuerle is with the Robert Bosch GmbH, D-72762 Reutlingen, Germany.}
\thanks{A. Steimer is with the Bosch Center for Artificial Intelligence (BCAI), Robert Bosch GmbH, D-71272 Renningen, Germany.}
}

%
%

\markboth{}%
{Shell \MakeLowercase{\textit{et al.}}: Bare Demo of IEEEtran.cls for IEEE Journals}
%



\maketitle

\begin{abstract}
Thermal management of power electronics and Electronic Control Units is crucial in times of increasing power densities and limited assembly space.
Electric and autonomous vehicles are a prominent application field.
Thermal Interface Materials are used to transfer heat from a semiconductor to a heatsink.
They are applied along a dispense path onto the semiconductor and spread over its entire surface once the heatsink is joined.
To plan this application path, design engineers typically perform an iterative trial-and-error procedure of elaborate simulations and manual experiments.
We propose a fully automated optimization approach, which clearly outperforms the current manual path planning and respects all relevant manufacturing constraints.
An optimum dispense path increases the reliability of the thermal interface and makes the manufacturing more sustainable by reducing material waste.
We show results on multiple real products from automotive series production, including an experimental validation on actual series manufacturing equipment.
\end{abstract}


%
\IEEEpeerreviewmaketitle

\section{Introduction}
%
%
%
%
\IEEEPARstart{T}{hermal} Interface Material (TIM) is used to transfer heat from a semiconductor to a heatsink.
During electronic packaging, TIM is applied onto the semiconductor and spreads over the cooling surface once the heatsink is joined.
This is valid for both for small Electronic Control Units (ECUs) and for larger power electronics modules such as inverters or chargers.
These products play a central role for electric and autonomous vehicles.
As power ratings increase and assembly space is limited, the thermal performance of the electronic package is a critical factor in the competitive market.

The demands regarding thermal performance for the entire electronic package translate to specific requirements regarding the TIM layer as follows.
To ensure an efficient heat flow, a high area coverage percentage of the cooling surface with TIM is desired.
A special case of a low area coverage is the formation of air entrapments or voids during the joining process of the heatsink.
For example, if TIM is applied in a circular shape onto the heatsink, air will be trapped inside and form a void area.
The effect of voids on the thermal performance can be \textit{devastating}~\cite{gowda_voids_2004}.
In automotive industry, the application of TIM onto to the semiconductor is typically carried out with selective dispensing technologies.
Prevalent technologies are time-pressure dispensing, rotary screw dispensing and positive displacement piston dispensing.
A high degree of automation is achieved by combining the dispensing system with a Computerized Numerical Control (CNC) machine.
The dispensing toolhead is programmed to travel along a predefined path.
While traveling along this path, the dispensing pump or piston can be (de-)activated as needed.
This results in a desired TIM dispensing pattern, i.e. a distribution of TIM material on the cooling surface.
This manufacturing process introduces additional requirements, which need to be considered during the design of the dispensing path: A) A low cycle-time is desired to maximize the number of manufactured parts per timeframe.
B) The amount of material dispensed per timeframe, i.e. the dispense rate, may not exceed a minimum and a maximum rate.
This is dependent e.g. on the material viscosity or the dispensing nozzle size.
C) The individual beads of the dispensing pattern should be separated from each other.
This enables an easier quality inspection and mitigates disturbances during the dispensing process.
D) Excess material flowing beyond the cooling surface should be minimized.
Excess material introduces unnecessary material cost.
E) Excess material can further violate sensitive areas - so-called taboo zones - in the vicinity of the cooling surface.
Those are e.g. screw holes or other electronic components.
This case is a violation of functional requirements and needs to be prevented.

The dispensing pattern is typically designed manually by an engineer.
The engineer designs the dispense path based on experience.
The design process is carried out in an iterative trial-and-error cycle.
It is supported by elaborate Computational Fluid Dynamics (CFD) simulations or by new heuristics~(\cite{baeuerle_rapid_2024, kaufmann_how_2023}).
Furthermore, manual experiments are carried out.
Restrictions regarding development time together with the complexity of the planning problem typically lead to resulting dispense patterns, which are sub-optimal with regard to above requirements. 

\section{Related work}
Traditional path planning aims to find a path, which connects a start point with a goal point while avoiding obstacles.
Well-known algorithms in this field are e.g. A*~\cite{hart_formal_1968} and Dijkstra~\cite{dijkstra_note_1959}.
Coverage Path Planning (CPP) aims to find a path \textit{that passes over all points of an area or volume of interest while avoiding obstacles}~\cite{galceran_survey_2013}.
It relates to the Traveling Salesman Problem (TSP), which aims to find the shortest route for visiting multiple cities.
TSP is a NP-hard problem to solve and can be efficiently approached with approximation algorithms.
A well-known example is the algorithm of Christofides and Serdyukov. 
The problem of dispense pattern design for TIM closely fits the problem formulation of CPP, but introduces an additional aspect: the area coverage depends on the complex material flow during the joining process.
This is in contrast to many related CPP applications, where the area within a fixed distance to the traveled path is being covered.

The CPP problem has been studied for many years.
Choset~\cite{choset_coverage_2001}, Galceran and Carreras~\cite{galceran_survey_2013} and Tan et al.~\cite{tan_comprehensive_2021} provide an overview over approaches to the CPP problem.
Choset and Pignon~\cite{choset_coverage_1998} proposed the Boustrophedon decomposition.
This decomposition splits the target area into cells, which can be covered by a back-and-forth motion.
Zelinsky et al.~\cite{zelinsky_planning_1993} use both the distance to the goal and the distance to the next obstacle to compute a numeric value for each grid cell.
This transformation is used as a basis to guide a robot through the target area.
Gabriely and Rimon~(\cite{gabriely_spanning-tree_2001, gabriely_spiral-stc_2002}) introduced an algorithm called Spiral-STC.
They compute a minimum spanning tree for a discretized target area.
The coverage path is created by travelling along the spanning tree.
Luo and Yang~(\cite{luo_solution_2002, yang_neural_2004, luo_bioinspired_2008}) propose a CPP approach that relies on the activation of a neural network.
The information about obstacles and unclean areas is propagated through an activity landscape and either repulses or attracts the robot.
Soltero et al.~(\cite{soltero_generating_2012, soltero_decentralized_2014}) use gradient-based optimization to adjust the waypoints of one or more quadcopters.
Coverage is maximized by pushing the waypoints towards the centroids of Voronoi cells, while keeping neighboring waypoints close to each other.
Kiemel et al.~\cite{kiemel_paintrl_2019} present a framework to enable CPP for industrial spray painting tasks using Reinforcement Learning.
Picciarelli and Foresti~\cite{piciarelli_drone_2019} focus on CPP for Unmanned Aerial Vehicles (UAVs).
They use a double deep Q-network and use a relevance map with spatially defined coverage requirements as an input.
Theile et al.~\cite{theile_uav_2020} add a movement budget to take into account not only the coverage goal, but also power constraints.
They use a 2D input map, which is processed by convolutional layers.
Ellefsen et al.~\cite{ellefsen_multiobjective_2017} use Genetic Algorithms to perform a multi-objective coverage path optimization on a discretized space for Autonomous Underwater Vehicles.
They treat the coverage, energy usage and collision occurrence as an objective rather than a constraint by introducing penalty values.
This yields paths that may not achieve complete coverage, but represent a good trade-off between all targets.
Batista and Zampirolli~\cite{batista_optimising_2019} discretize the target area and use a Genetic Algorithm to plan a coverage path for a tool cleaner.
They allow repeated visiting of both path edges and path vertices.
They further allow both open and closed paths.
Yakoubi and Laskri~\cite{yakoubi_path_2016} split the coverage path of a vacuum cleaner robot into short sub-paths.
They optimize the combination of the sub-paths using Genetic Algorithms.
Popovic et al.~\cite{popovic_online_2017} optimize the area coverage path of an UAV surveillance task while considering constraints such as a time budget.
They assume a pyramid-like field of view of a camera mounted on a UAV.
The planning procedure is split into two parts.
The first part is the selection of viewpoints in 3D space.
In a second step, a continuous trajectory is defined based on the chosen viewpoints using the optimization algorithm Covariance Matrix Adaption Evolution Strategies (CMA-ES)~\cite{hansen_cma_2016}.
Strubel~\cite{strubel_coverage_2019} presents a Coverage Path Planning approach, which is suitable for an area surveillance task using a UAV.
He divides the problem into two separate tasks: waypoint finding and connecting the waypoints to form a path.
For the waypoint finding, the assumption of a rectangular field of view of a camera mounted to an UAV is used.
Thus, the area coverage is composed of a multitude of individual rectangles.
The rectangles are allowed to overlap each other.
Coverage Path Planning has also been studied in the context of Additive Manufacturing (AM)~(\cite{kulkarni_review_2000, oropallo_ten_2016}).
Some approaches mainly rely on back-and-forth patterns (e.g.~\cite{jin_adaptive_2011}), optimize velocity profiles rather than the global path pattern (e.g.~\cite{roveda_pairwise_2021}) or adjust the path pattern only locally (e.g.~\cite{comminal_motion_2019}).
The workspace can be processed with a medial axis transform~\cite{blum_transformation_1967}, which can be used as a basis for path planning with a variable path width~\cite{kao_optimal_1998}.
Using a variable path width in AM has been further studied, e.g. by Ding et al.~\cite{ding_adaptive_2016}, Wang et al.~\cite{wang_variable_2019}, Xiong et al.~\cite{xiong_process_2019} and Hornus et al.~\cite{hornus_variable-width_2020}.
Xiong et al. further perform an optimization of process parameters such as material feed rate to achieve a desired height and width of the deposited material.
In order to perform this optimization, they use Gaussian Processes to model the relationship between in- and outputs.
This data-driven model is efficient in terms of computing time and accuracy.
Similarly, Ding et al.~\cite{ding_towards_2016} use an Artificial Neural Network (ANN) instead of Gaussian Processes, since ANNs are well-suited to model complex relationships.
Advantages of printing with a variable path width generally include prevention of void areas, increased production speed and a higher geometrical accuracy~\cite{xiong_process_2019}.
Additional criteria to be studied in this setting are e.g. the influence on residual stress~\cite{xiong_process_2019} and possible voids between individual printing planes~\cite{hornus_variable-width_2020}.
Flaig et al.~\cite{flaig_how_2023} propose to invert a recently introduced flow behavior model for adhesives~\cite{kaufmann_how_2023}.
This inversion results in dispense patterns, which cover the target area.
This has been tested with manual experiments~\cite{flaig_practical_2023}.
The resulting patterns exhibit a branched shape rather than a continuous path.
This is similar to the result of a medial axis transform, which has been used for path planning in additive manufacturing before~\cite{kao_optimal_1998}.
The branched shape of the resulting patterns would require to an automated dispensing system to travel multiple times over the same edge.
Assuming a constant dispense rate, this will make it difficult to match the proposed material distribution in practice.
Furthermore, the beads are not separated from each other.
This yields to difficulties during quality inspection.
Flaig et al. themselves state that the resulting patterns \textit{may not yet be well suited for practical application, e.g. by automatized dosing systems}~\cite{flaig_how_2023}.
A more recent publication of the same research group exhibits similar issues with the resulting patterns~\cite{kaufmann_optimized_2024}.
Due to the failure to comply with technical constraints, these approaches cannot be used for automated dispensing systems which are used for automotive series manufacturing.

Another relevant field of research is the optimization of trajectories in the field of robotics.
It typically involves not only planning the path in terms of a path through space, but also the speed or even the acceleration along this path.
Gradient-based optimizers can be utilized for such problems~\cite{bryson_applied_1975}.
However, gradient-based optimizers strongly depend on the initial trajectory guess and may get stuck in a local minimum.
This corresponds to a non-optimal solution.

\begin{figure}[hbt]
	\centering
	\includegraphics[width=0.9\columnwidth]{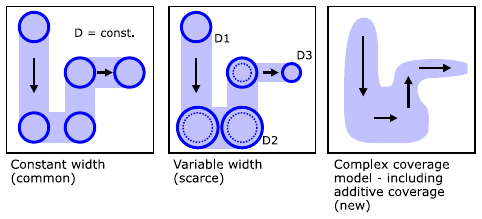}
	\caption{Area coverage types during Coverage Path Planning problem settings. Most approaches consider a constant path width.}\label{fig:01_cpptypes}
\end{figure}

The above approaches can be clustered according to the underlying area coverage behavior for the cases, which are shown in Figure~\ref{fig:01_cpptypes}.
Almost all of the classic coverage path planning approaches consider a constant path width as shown on the left of Figure~\ref{fig:01_cpptypes}. A constant width reflects the nature of many of the relevant physical systems accurately.
Examples are lawn mowers, milling machines, cleaning robots, 3D printers or harvesting machines.
Very few approaches consider a varying path width as shown in the center of Figure~\ref{fig:01_cpptypes}. During additive manufacturing, the bead size can be slightly modified.
However, this variation of the path width is bound to narrow limits, since increasing the bead width simultaneously influences also the bead height.
The bead height is generally desired to be constant within a printed layer.
Furthermore, the bead width of the printed materials is rather limited.
Applying the approaches from additive manufacturing to dispensing of TIM is not feasible, since air entrapments would form for the resulting patterns.

Furthermore, a few approaches with varying path width came up for UAVs.
They consider a pyramid-like field of view leading to a rectangular coverage area on the target surface for an individual viewpoint.
These approaches~(\cite{popovic_online_2017, strubel_coverage_2019}) are limited to a rectangular coverage for a single viewpoint.
For many practical use-cases, the horizontal dimensions are proportionally much larger than the vertical dimension.
In this case, the approach will become similar to a constant path width.
The nature of TIM dispensing systems is fundamentally different.
For example, air entrapments are naturally not considered for UAVs at all.
The coverage path planning problem for TIM has not only a varying path width, but a complex underlying flow behavior model as shown on the right of Figure~\ref{fig:01_cpptypes}.
A core difference to many related CPP approaches is that the TIM material will add up - and respectively increase the covered area - if the same spatial area is covered twice.
This kind of complex coverage model is missing in almost all related works.
An automated dispense pattern optimization based on such a flow behavior model has first been shortly mentioned in a pre-print~\cite{baeuerle_rapid_2022} of Baeuerle et al.~\cite{baeuerle_rapid_2024}.
In parallel to our research activity, the research group of Flaig and Kaufmann et al. (\cite{flaig_how_2023, kaufmann_optimized_2024}) has also tried to optimize the area coverage based on a complex coverage model.
The limitations of their approach have been outlined above.

To our knowledge, no further approaches considering a complex coverage behavior for the CPP problem have been proposed.
Up to now, a feasible solution for this kind of problem is still missing.
We propose an automated dispense pattern design approach, which yields a dispense pattern for a given cooling surface geometry.
It considers the requirements, which are relevant for the dispensing equipment used in series manufacturing.
We elaborate the detailed setup of our approach and the influence of particular parameters on the quality of the resulting dispense patterns.
We show results for real automotive products.
We further benchmark our automated method against manually designed dispense patterns, which have been proposed in past by design engineers for regular products of a major automotive company.

Section 2 outlines our new CPP approach for TIM.
Section 3 describes the evaluations, which we carry out.
This includes a study of the optimization setup and the benchmarking of specific target product geometries.
Section 4 contains the respective results.
Section 5 makes both advantages and disadvantages transparent and highlights the key features, which we found to be relevant for a successful path planning.
Section 6 concludes our work.

\section{Methodology}
In this section, we give an overview of our optimization setup.
Subsection 1 contains the utilized optimizer along with the underlying model for the dispensing and packaging processes.
Subsection 2 explains the detailed setup of the objective function.

\subsection{Optimizer and process model}

\begin{figure}[hbt]
	\centering
	\includegraphics[width=0.9\columnwidth]{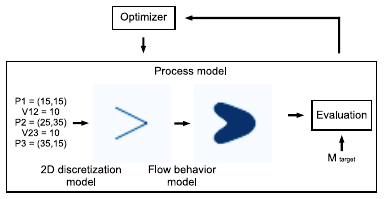}
	\caption{Overall approach: optimizer interacting with process model}\label{fig:02_overall_pipeline}
\end{figure}
%
%
%
%
%
%
%
%
%

Baeuerle et al.~\cite{baeuerle_rapid_2024} have presented a model for both for the dispensing and the packaging processes.
The input of the first model is the dispense path.
This input path is shown on the left side of the process model in Figure~\ref{fig:02_overall_pipeline}.
It is parameterized with a series of point coordinates, which are connected by lines.
For each line, one parameter corresponding to the respective TIM material feedrate is set.
The lines form a polygonal chain, which is transferred to a two-dimensional representation with a 2D discretization model.
This represents the dispensed state of material.
Two flow behavior models are introduced, which model the flow of material during the packaging process, i.e. during pressing the heatsink onto the dispensed TIM.
The output is the compressed state of TIM.
The first flow behavior model is a heuristic based on the material volume conservation and does not include any further effects such as viscosity or friction.
The second flow behavior model is an ANN trained on the data of the heuristic.
Both models were validated on experimental data.
The work was mainly focused on the flow behavior models and their computation speed.
Both the material distribution after dispensing and the material distribution after joining are discretized to $50\,\times\,50$ grid cells.
A number in each cell represents the respective TIM amount per grid cell.~\cite{baeuerle_rapid_2024}

Our current work is built on top of the 2D discretization model and the (heuristic) flow behavior model of Baeuerle et al.~\cite{baeuerle_rapid_2024}.
The overall approach is sketched in Figure~\ref{fig:02_overall_pipeline}, with both models being embedded into the process model.
Both the ANN and the heuristic are a flow behavior model as depicted in the center of Figure~\ref{fig:02_overall_pipeline}.
While the ANN is quicker in terms of computation speed, it introduces a small error with respect to the predicted compressed material distribution.
In order to analyze the performance of the optimization as accurately as possible within this work, we forego the advantages of the ANN with regard to computation speed and instead work with the original heuristic to retain the best accuracy.
We use the 2D discretization model and the heuristic flow behavior model to evaluate a given dispense path and extend it with several evaluation strategies.
The evaluation outputs are aggregated to a single objective function.
This objective function serves as a dependent variable for the optimizer.
The independent variables to be adjusted by the optimizer are the point coordinates of the dispense path.
While the feedrate could be set individually for each path segment, we keep it constant along the entire dispense path.
The feedrate is computed to match the total amount of TIM, which is needed to cover a given surface geometry.
For the optimization itself, we utilize the optimizer CMA-ES as it is implemented in the \textit{Optuna} library~\cite{akiba_optuna_2019}.
CMA-ES can efficiently walk through a large search space.
It considers the correlations of the parameters.
CMA-ES has shown a very good performance on black box optimization problems~\cite{hansen_comparing_2010}.

\subsection{Shaping of objective function}
\begin{figure*}[hbt]
	\centering
	\includegraphics[width=1.9\columnwidth]{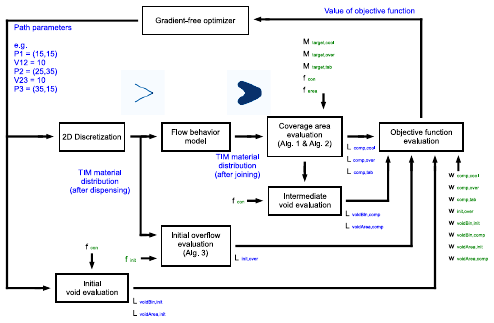}
	\caption{Detailed workflow with intermediate in- and outputs}\label{fig:02_detailed_pipeline}
\end{figure*}

The objective function is comprised of several components as shown in Figure~\ref{fig:02_detailed_pipeline}.
They can be summarized to the following groups: Coverage area evaluation, initial overflow evaluation and void evaluation.
These evaluation strategies are integrated into the above model of the physical manufacturing process.
The following paragraphs outline the detailed computation of each loss component from those groups.

We apply weighting functions within the individual components using the following identifiers:
\begin{multline}\label{eq:weightingfunctions}
\begin{gathered}
\text{none}: f(x)=0 \\
\text{con}: f(x)=x \\
\text{lin}: f(x)=ax \\
\text{squ}: f(x)\,=\,x^2 \\
~~~~~~~~~~~~~~\text{log}: f(x)\,=\,-1\cdot\log(1-x).~~~~~~~~~~~~~~
\end{gathered}
\end{multline}

\subsubsection{Coverage area evaluation}

\begin{figure}[hbt]
	\centering
	\includegraphics[width=.8\columnwidth]{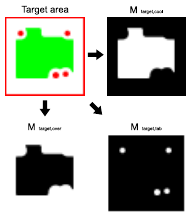}
	\caption{Split of overall target area in three individual target area types. $M_{target,cool}$ (green color) defines the cooling area, $M_{target,over}$ (white and red color) the non-cooling area and $M_{target,tab}$ (red color) the taboo zones. In each greyscale image, the white color hue indicates the respective spatial location belonging to the specified target area type.}\label{fig:02_target_area_types}
\end{figure}

The coverage area evaluation is the most obvious component of our objective function.
We compare the actual coverage with the desired coverage.
We distinguish three area types in our desired coverage as shown in Figure~\ref{fig:02_target_area_types}.
The green color represents the cooling surface, which should be covered.
Its area definition is stored in the matrix $M_{target,cool}$, which can be interpreted as a greyscale image with values in the range from zero to one.
The white color represents the surface of surrounding parts, where the TIM may overflow into, but which would increase material waste.
This surface is stored in the matrix $M_{target,over}$.
The red color represents so-called taboo zones, which must not be touched by TIM.
These are stored in $M_{target,tab}$.

A reduced color saturation represents the grid cell having several area assignments in proportion to the saturation.
I.e. a cell with a light-green color hue should ideally be covered only proportionally to its color saturation.
A cell with a light-red color hue is weighted less than a fully red cell during evaluation.
Since we work with a $50\,\times\,50$ grid resolution, the resolution is limited to the discretized cells.
When working without the reduced color saturation, the accuracy would be bound to the discretization, i.e. a cell can be either fully green, white or red.
Allowing a reduced color saturation along with several area assignments/colors enables us to represent an area boundary, which ends between two grid cells.
Effectively, this enables a higher resolution for the area definitions.

We test different evaluation strategies for the material coverage.
The first strategy \textit{S-con} simply sums up the material amounts in each area type and applies either no weighting (\textit{con}), square weighting (\textit{squ}) or logarithmic weighting (\textit{log}).
The second strategy \textit{S-area} applies the weighting on the area itself already, i.e. before summing up the individual values.
For this kind of weighting we utilize linear (\textit{lin}), square (\textit{squ}) or logarithmic (\textit{log}) weighting.

\begin{algorithm}[!ht]
	\caption{S-con}\label{alg:ConstantAreaDistanceWeighting}
	\begin{algorithmic}[1]
		\STATE \hspace{0.5cm}S-con (M\_comp, M\_target, M\_target,cool, f\_con)
		\STATE \hspace{0.5cm}targetCover = multiply(M\_comp, M\_target)
		\STATE \hspace{0.5cm}normCover = targetCover / sum(M\_target,cool)
		\STATE \hspace{0.5cm}clipCover =min(normCover, 1)
		\STATE \hspace{0.5cm}If M\_target = “M\_target,cool”:
		\STATE \hspace{0.5cm}~~L\_comp = f\_con(1-clipCover)
		\STATE \hspace{0.5cm}Else:
		\STATE \hspace{0.5cm}~~L\_comp = f\_con(clipCover)
		\STATE \hspace{0.5cm}Return L\_comp
	\end{algorithmic}
\end{algorithm}

The strategy \textit{S-con} is calculated according to the pseudocode in Algorithm~\ref{alg:ConstantAreaDistanceWeighting}.
The function has the following inputs: $M_{comp}$, $M_{target}$, $M_{target,cool}$, $f_{con}$.
$M_{comp}$ is a matrix with float values within the interval from zero to one.
It represents the actual compressed state.
$M_{target}$ is a matrix with integer values of either zero or one.
It can be one of the predefined target states $M_{target,cool}$, $M_{target,over}$ or $M_{target,tab}$, which have been introduced above.
$f_{con}$ is a string and specifies the weighting function to be used as defined in Equation~\ref{eq:weightingfunctions}.
Such a weighting function affects the shape of the overall objective function and can have an influence on the convergence behavior during the optimization.
For example, the objective function might exhibit a plateau, i.e. a region in its input parameter space for which its output values do not change significantly.
Here, using such a weighting function would turn this flat plateau into an inclining region and enable an optimization algorithm to identify a better search direction.
Algorithm~\ref{alg:ConstantAreaDistanceWeighting} returns the float value $L_{comp}$.
Depending on the given $M_{target}$, this is the corresponding objective function term for each of the area types $L_{comp,cool}$, $L_{comp,over}$ or $L_{comp,tab}$.
Algorithm~\ref{alg:ConstantAreaDistanceWeighting} needs to be executed first for $M_{target,cool}$.
This will store a normalization value in a global variable, which is used for the remaining target area types.
For each area type, the respective greyscale image ($M_{target,cool}$, $M_{target,over}$ or $M_{target,tab}$) is multiplied cell-wise with the material distribution after the joining process $M_{comp}$ and summed up.
This yields one actual coverage sum per area type.
This sum is then normalized with the sum of a fully covered cooling area and clipped to a maximum equal to one.
Dependent on the desired weighting function, the normalized area sum is then squared, processed with a logarithmic function or left as-is (no weighting).
The results for each target area type are weighted relatively to each other by multiplying them with another weighting factor $w_{comp}$:
\begin{multline}\label{eq:lcomp1}
\begin{gathered}
~~~~~~~~L_{comp,total}~=~w_{comp,cool}~\cdot~L_{comp,cool}~~~~~~~~ \\
~~~~~~~~~~~~~~~~+~w_{comp,over}~\cdot~L_{comp,over} \\
~~~~~~~~~~~~~~~+~w_{comp,tab}~\cdot~L_{comp,tab},
\end{gathered}
\end{multline}
with $w_{comp,cool}$ being the weighting factor for the cooling surface, $w_{comp,over}$ being the weighting factor for the overflow area and $w_{comp,tab}$ being the weighting factor for the taboo zones.

\begin{algorithm}[!ht]
	\caption{S-area}\label{alg:NonlinearAreaDistanceWeighting}
	\begin{algorithmic}[1]
		\STATE \hspace{0.5cm}S-area (M\_comp, M\_target, M\_target,cool, f\_area)
		\STATE \hspace{0.5cm}distTrafo = cv2.distanceTrafo(M\_target)
		\STATE \hspace{0.5cm}If M\_target = M\_target,cool:
		\STATE \hspace{0.5cm}~~Global maxDistCool = max(distTrafo)
		\STATE \hspace{0.5cm}clipDistTrafo = clip(distTrafo, [0, maxDistCool])
		\STATE \hspace{0.5cm}normDistTrafo = clipDistTrafo / maxDistCool
		\STATE \hspace{0.5cm}fctDistTrafo = f\_area(normDistTrafo)
		\STATE \hspace{0.5cm}binaryCover = where(M\_comp \textgreater~0, 1, 0)
		\STATE \hspace{0.5cm}If M\_target = M\_target,cool:
		\STATE \hspace{0.5cm}~~Global maxMatSumCool = sum(fctDistTrafo)
		\STATE \hspace{0.5cm}wghtCover = multiply(binaryCover, fctDistTrafo)
		\STATE \hspace{0.5cm}normCover = sum(wghtCover) / maxMatSumCool
		\STATE \hspace{0.5cm}L\_comp = clip(normCover, [0, 1])
		\STATE \hspace{0.5cm}Return L\_comp
	\end{algorithmic}
\end{algorithm}

\begin{figure}[hbt]
	\centering
	\includegraphics[width=0.8\columnwidth]{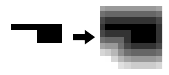}
	\caption{Spatial weighting of target area. The weighting factors are specified for each grid cell. They increase proportionally to the distance between the respective grid cell and the reference area.}\label{fig:02_area_weighting}
\end{figure}

The strategy \textit{S-area} is calculated according to the pseudocode in Algorithm~\ref{alg:NonlinearAreaDistanceWeighting}.
It is an alternative to the strategy \textit{S-con}.
Its input values, identifiers and output values are the same, i.e. Equation~\ref{eq:lcomp1} is valid here as well.
In contrast to \textit{S-con}, the weighting is applied at the individual grid cells based on the spatial location thereof.
The greyscale image for each area type is processed with the function \textit{distanceTransform()} from the \textit{OpenCV} library~\cite{bradski_opencv_2000} using an Euclidean distance metric.
An example of a transformed area with a reduced resolution of $10\,\times\,10$ grid cells is shown on the right of Figure~\ref{fig:02_area_weighting}.
Its grey scale values correspond to the weighting factors for each grid cell.
Grid cells close to the target area in the center exhibit low values (dark grey).
The weights increase proportionally to the distance to the target area, i.e. outer regions have a maximum weight (white color).
In the case of the overflow or taboo area, its values are clipped to the same maximum distance-transformed value as calculated for the cooling area.
All resulting cell values are normalized to a range from zero to one.
For each grid cell, the values are weighted using the linear, square or logarithmic function or left as-is in the case of no cell-wise weighting.
The result is cell-wise multiplied with the actual coverage of TIM after joining.
The sum is then normalized again using the sum of the cooling area and clipped to a range from zero to one.
The results for each target area type are weighted relatively to each other in analogy to Equation~\ref{eq:lcomp1}.

Two hyperparameters are set to control the weighting of the area loss.
The parameter $f_{area}$ is set to be either \textit{con}, \textit{lin}, \textit{squ} or \textit{log}.
The parameter $f_{con}$ is set to be either \textit{con} or \textit{log}.
Setting $f_{area}$=\textit{con} will set $L_{comp}$ to the result of \textit{S-con()} as defined in Algorithm~\ref{alg:ConstantAreaDistanceWeighting} (\textit{S-area}() will not be called).
Setting $f_{area}$ to either \textit{lin}, \textit{squ} or \textit{log} will compute $L_{comp}$ for the respective target area type using the result of \textit{S-area}() as defined in Algorithm~\ref{alg:NonlinearAreaDistanceWeighting} (\textit{S-con}() will not be called).
Thus, only one of the weightings can be active at the same time, i.e. $L_{comp}$ as defined in Equation~\ref{eq:lcomp1}  will take the value of only one of the outputs of either \textit{S-con()} or \textit{S-area()}.
\textit{S-con} and \textit{S-area} utilize the same relative weighting factors $w_{comp}$ and return the resulting loss term with the same identifier $L_{comp}$.

\subsubsection{Initial overflow evaluation}
Our optimization problem is set up unconstrained, i.e. we do not pose explicit constraints on the independent variables.
Thus, the optimizer may set path points beyond the cooling surface.
To prevent this, we formulate an additional term, which is added to the objective function.
We utilize part of the computation approach presented during the coverage area evaluation as described in pseudocode in Algorithm~\ref{alg:InitOverflowWeighting}.
The function has the following inputs: $M_{initial}$, $M_{target,over}$, $f_{init}$.
It works in analogy to Algorithm~\ref{alg:NonlinearAreaDistanceWeighting} and is calculated for the initially dispensed material (instead of the compressed material).
Here, the spatial distance values for the individual grid cells are clipped to a maximum distance of half of the entire grid size (maxDist=25).
Similar to $f_{area}$, $f_{init}$ is set to be either \textit{lin} or \textit{log} as defined in Equation~\ref{eq:weightingfunctions}.
We refer to the output of this computation with $L_{init,over}$.

\begin{algorithm}[!t]
	\caption{S-init}\label{alg:InitOverflowWeighting}
	\begin{algorithmic}[1]
\STATE \hspace{0.5cm}S-init (M\_{initial}, M\_{target,over}, f\_init)
\STATE \hspace{0.5cm}distTrafo = cv2.distanceTrafo(M\_target,over)
\STATE \hspace{0.5cm}maxDist = 25  \# Half of the overall grid height/width
\STATE \hspace{0.5cm}clipDistTrafo = clip(distTrafo, [0, maxDist])
\STATE \hspace{0.5cm}normDistTrafo = clipDistTrafo / maxDist
\STATE \hspace{0.5cm}fctDistTrafo = f\_init(normDistTrafo)
\STATE \hspace{0.5cm}binaryCover = where(M\_ initial \textgreater 0, 1, 0)
\STATE \hspace{0.5cm}wghtCover = multiply(binaryCover, fctDistTrafo)
\STATE \hspace{0.5cm}maxMatSum = sum(fctDistTrafo)
\STATE \hspace{0.5cm}L\_ init,over = sum(wghtCover) / maxMatSum
\STATE \hspace{0.5cm}Return L\_ init,over
	\end{algorithmic}
\end{algorithm}

\subsubsection{Initial and intermediate void evaluation}

\begin{figure}[hbt]
	\centering
	\includegraphics[width=0.9\columnwidth]{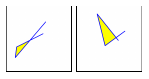}
	\caption{Two exemplary TIM dispense patterns with the material itself depicted in  blue color. The highlighted areas (yellow) indicate initial voids as they may appear directly after the dispensing process.}\label{fig:02_initial_void}
\end{figure}
%

Voids can be prevalent directly after the dispensing.
If the path segments cross each other, this will yield a direct formation of a void.
This case is easy to detect, since it is obvious from the dispense path.
We call this kind of voids \textit{initial voids}.
In order to detect them with an algorithm, we scale up the resolution from $50\,\times\,50$ to $1000\,\times\,1000$ and re-apply the discretization.
We then apply the \textit{OpenCV}~\cite{bradski_opencv_2000} function \textit{cv2.connectedComponentsWithStats()} to detect the number of connected areas and to measure the size of the areas.
If the number of connected areas is larger than two, a void is present.
The detection of an initial void is shown in Figure~\ref{fig:02_initial_void}.
We assess two types of integrating this kind of void into our objective function.
First, we define the binary term $L_{voidBin,init}$ to be either 0 (in case of no void) or 1 (in case of at least one void).
Second, we describe the computation of the term $L_{voidArea,init}$.
The area \textit{voidArea} is equal to the area of the void as computed by \textit{cv2.connectedComponentsWithStats()}.
In case of multiple voids, the areas are summed up.
We divide \textit{voidArea} by the sum of $M_{target,cool}$ and assign the result to $L_{voidArea,init,norm}$.
We clip the value range to a maximum of one and apply the weighting function defined by $f\_{con}$.
The result is identified by $L_{voidArea,init}$.

\begin{figure}[hbt]
	\centering
	\includegraphics[width=0.5\columnwidth]{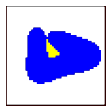}
	\caption{An exemplary TIM dispense pattern. The shown state is captured halfway through the joining process. The material (blue) is spreading, but has not reached its final state yet. The highlighted area (yellow) indicates an intermediate void, which can occur during the joining process.}\label{fig:02_intermediate_void}
\end{figure}

Voids can also form during the joining process.
This is the case, when the TIM fluid fronts collide in a way that traps air within.
Especially for complex dispense paths, this may not be obvious at first sight and we refer to this kind of void as an \textit{intermediate void}.
In order to detect this kind of voids, we leverage the iterative nature of the heuristic flow behavior model of Baeuerle et al.~\cite{baeuerle_rapid_2024}
Since the compression of the material is computed at discretized gap heights, which decrease iteratively, we can apply the function \textit{cv2.connectedComponentsWithStats()} at each iteration.
This enables us to detect most of the intermediate voids.
The presence of these voids is captured by the binary term $L_{voidBin,med}$, which is defined in analogy to $L_{voidBin,init}$.
We apply the same procedure of measuring and normalizing the void area as with the initial voids and obtain $L_{voidArea,med}$.
The void area is calculated at the iteration of the flow behavior model, where the intermediate void is detected first.
If the void is present during subsequent iterations of the flow behavior model, the areas are not aggregated.
A visual depiction of an intermediate void is shown in Figure~\ref{fig:02_intermediate_void}.

\subsubsection{Objective function evaluation}
We obtain the overall objective function by summing up the above terms.
Each term is weighted with a relative weighting factor $w$:
\begin{multline}
\begin{gathered}
~~~~~L~=~w_{comp,cool}~\cdot~L_{comp,cool}~~~~~ \\
~~~~~~~~+~w_{comp,over}~\cdot~L_{comp,over} \\
~~~~~~+~w_{comp,tab}~\cdot~L_{comp,tab} \\
~~~~~~+~w_{init,over}~\cdot~L_{init,over} \\
~~~~~~~~+~w_{voidBin}~\cdot~L_{voidBin,init} \\
~~~~~~~~+~w_{voidBin}~\cdot~L_{voidBin,med} \\
~~~~~~~~~+~w_{voidArea}~\cdot~L_{voidArea,init} \\
~~~~~~~~~~+~w_{voidArea}~\cdot~L_{voidArea,med}.
\end{gathered}
\end{multline}

Note that the individual components $L_{comp}$ and $L_{voidArea}$ depend on the value of $f\_{con}$.
$L_{comp}$ further depends on $f_{area}$.
$L_{init,over}$ depends on $f_{init}$.

\section{Experimental Setup}
This section describes the setup of our analysis in order to evaluate our proposed optimization approach.

\subsection{Evaluation on products A-D (existing products)}
\begin{figure}[hbt]
	\centering
	\includegraphics[width=0.9\columnwidth]{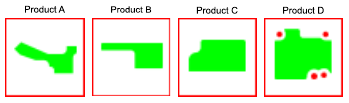}
	\caption{Target area definition of four real-world automotive products. These are used during the benchmarking of our proposed optimization approach against expert paths.}\label{fig:03_existing_products}
\end{figure}

We evaluate our new optimization approach on four existing automotive products from the fields of electronic control units and power electronics.
The definition of the desired target coverage is shown in Figure~\ref{fig:03_existing_products}.
The color coding follows the area definitions above: Green is the cooling surface, white is the non-critical overflow area and red is the taboo zone area.
We run the stochastic optimization multiple times for several days.
This yields many result paths for each product.
The results are sorted by the area coverage of the cooling surface and a suitable path is chosen manually from the top of the list.
Since those products are being manufactured already, we can acquire the corresponding expert path, i.e. the dispense path which has been designed by an engineer.
At that time, the optimization was not available to support the manual planning process.
We transfer the expert path into a representation which is compatible to our parameterization of the path.
This enables a direct comparison of the optimization result with the expert path using the flow behavior model proposed by Baeuerle et al.~\cite{baeuerle_rapid_2024}
The transfer of the expert path into a compatible parameterization is carried out as follows.
The coordinates can be transferred directly from the coordinates of the technical drawing.
Some products exhibit a rather large mechanical tolerance with respect to the gap height after joining.
This makes it difficult to set the dispense amounts correctly.
To enable a fair comparison, we adjust the dispense amounts of the expert path in a way that yields an almost complete coverage of the cooling surface.
We notice that at some point further increasing the dispense amounts leads to a disproportionally high material overflow.
We stop adjusting the dispense amounts before exceeding this point.
Thus, not all expert paths yield a 100\,\% coverage of the cooling surface.
This coverage is reported as \textit{coverage of cooling surface} in Table~\ref{tab:04_optvshuman}.
We set the dispense amounts of the optimized path in way that yield the very some coverage percentage of the cooling surface.
At this exact coverage percentage, we compare the optimized path against the expert path.
For both the expert path and the optimized path, we measure the amount of TIM that was overflowing beyond the cooling surface and divide it by the amount of material within the cooling surface.
This is reported as overflow ratio in Table~\ref{tab:04_optvshuman}.

\subsection{Evaluation on product E (new product)}

\begin{figure}[hbt]
	\centering
	\includegraphics[width=0.7\columnwidth]{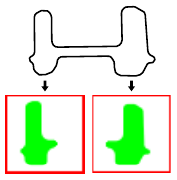}
	\caption{Test on product E, which is a new product. The cooling surface of product E is sketched at the top. It is split into two sub-areas as shown on the bottom, which are to be optimized individually. The connecting region in the center is not included in the optimization.}\label{fig:03_new_product}
\end{figure}
In parallel to our research activity, a new product was being introduced.
We offered to assist the manual dispense path design with our proposed optimization approach.
Due to the large cooling surface area, we split the area into smaller sub-areas and run the optimization on each sub-area separately.
To ensure a continuous dispense path for the aggregated area, we set the path points on the connecting edges to a fixed coordinate.
The product area and its split into the sub-areas are shown in Figure~\ref{fig:03_new_product}.
The proposals from our approach are slightly adjusted manually and used during the series ramp-up of the respective product.
For this product, TIM is applied onto the mounted electronic components.
The underlying flow behavior model assumes plane surfaces.
Thus, the real-world surface geometry formally violates the constraints of the flow behavior model.
Even in the case of such a violation, the flow behavior model can be utilized to enable an optimization of the dispense path.

\subsection{Configuration of objective function}
We have proposed multiple components of the objective function, which can be weighted and then combined in many different ways.
This can be interpreted as a hyperparameter optimization problem in itself.
Due to the stochastic nature of the optimizer, results vary between multiple optimization runs.
Different cooling surface geometries might have different sets of optimal hyperparameters.
We manually define a set of hyperparameters to be assessed for a single product in the scope of this work.
We are aware, that this might not constitute an overall optimum and hope to gain insights for future studies.
We formulate the optimization as a minimization problem.
In the best case, the cooling surface is covered by 100\,\%.
We set the feedrate to be constant along the entire path and the total initial TIM amount equal to the theoretically needed TIM amount to cover exactly 100\,\% of the cooling surface.
In this setting, $w_{comp,cool}$ can be set to zero: this overall fixed TIM amount combined with a penalty for the overflowing material will yield an optimum value for a complete coverage of the cooling surface.
An alternative to this setup would be to treat the overall TIM amount as an independent variable (to be optimized) along with setting $w_{comp,cool}$ to a negative value.
As the values of the remaining objective function terms are essentially to be optimized in relation to each other and scaling all weights with the same factor will have no effect, we set $w_{comp,over}$\,=\,1.
The clearance of a taboo zone constitutes a functional requirement for the resulting product.
A violation of a taboo zone leads to the product being sorted out during quality inspection.
If we estimate the cost of TIM to be around 0,50 Euro per part and the cost of an entire power electronics product to be around 500 Euros, we assume a weighting of $w_{comp,taboo}$\,=\,1000\,$\cdot$\,$w_{comp,over}$ to be a reasonable choice.
A similar estimation can be done for the voids.
Since dispensing TIM beyond the cooling surface obviously does not make any sense, we define the weight for $w_{init,over}$ to be in similar orders of magnitude as $w_{comp,taboo}$ and $w_{void}$.
In our trial we test a rather broad range of order of magnitudes of $w \in [10, 100, 1000, 10000]$ for each of $w_{comp,taboo}$, $w_{void}$ and $w_{init,over}$.
We further test different weighting functions on grid-cell level as described above by setting $f_{con}$, $f_{area}$ and $f_{init}$.
A full grid search of this space is not feasible.
The exact hyperparameter combinations which we assess is documented in the appendix.
Each hyperparameter configuration is executed 100 times for paths with a length of 5 to 10 segments and with 1000 optimization iterations of CMA-ES.
For each hyperparameter setting, we compute three metrics: the average coverage of the cooling surface, an auxiliary value for the convergence properties and the average of both, which we refer to as \textit{average performance ratio}.
The coverage is readily available from the coverage area evaluation procedure.
We calculate the average of the coverage percentage over all runs in each individual hyperparameter setting.
We obtain a single average coverage value in the range from zero to one.
The auxiliary value for the convergence properties is calculated as follows.
We define a binary value for each hyperparameter setting with a value of one corresponding to the convergence to a good local minimum.
A value of zero corresponds to a not usable dispense path.
The conditions are: at least 80\,\% coverage of the cooling surface area, a maximum violation of a taboo zone area of 1\,\% relative to the cooling surface area, a maximum void area of 5\,\% relative to the cooling surface area.
These binary values are added up and divided by the total number of optimization runs for each hyperparameter setting.
This again yields a single auxiliary value in the range from zero to one.

\subsection{Compliance with mechanical tolerances}
Real products are subject to mechanical tolerances.
With regard to the dispense path planning, especially the tolerances with respect to the final gap height after joining are relevant.
The tolerances are in some cases significant and the tolerance width can even be in the same range as the total gap height.
This has implications on the dispense path planning: In a maximum gap setting, the coverage of the cooling surface needs to be sufficient.
In a minimum gap setting, the overflowing material may not violate any taboo zone.
We comply with this kind of tolerance by evaluating the material coverage at the maximum gap height in a first step.
In a second step, we compress the TIM material further down to the minimum gap height and evaluate the material coverage again.
The objective terms with regard to area coverage are added up for both gap heights.
This forces the optimizer to find a dispense path, which both covers as much of the cooling surface as possible in the maximum gap setting and avoids the violation of taboo zones in a minimum gap setting.

\section{RESULTS}
In this section we present results from our proposed optimization approach: we present optimized dispense paths for real-world products and we evaluate the influence of the hyperparameters of the optimization.

\begin{figure}[hbt]
	\centering
	\includegraphics[width=1\columnwidth]{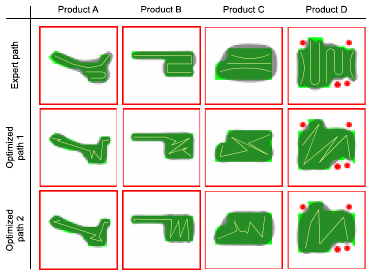}
	\caption{Comparison of dispense patterns which are either designed by experts or are a result from our proposed coverage path planning. The shown products are real products, which had been in series manufacturing before our approach was available. Thus, the expert paths are the result of the conventional dispense pattern design process under real-world conditions. They serve us as a realistic benchmark.}\label{fig:04_optvshuman}
\end{figure}

The optimization results are visualized along with the expert paths in Figure~\ref{fig:04_optvshuman}.
The dispense paths are shown in a light yellow cover.
The resulting TIM coverage is shown as an overlay in a grey color on top of the target area definitions introduced in the previous section.
Table~\ref{tab:04_optvshuman} shows the ratio of overflowing TIM material relative to TIM material amount within the cooling surface.
All of the resulting dispense patterns respect the functional constraints regarding both voids and taboo zones.
Manufacturing tolerances (see Subsection~\ref{sec:compliancemechanicaltolerances}) are not included in this table, i.e. we optimize for the nominal gap height.
The optimizer outperforms the expert paths in all tested products.
The advantage of using the optimization in terms of potential material savings is considerable.
This is especially the case for the shown products, since they are produced in high numbers during automotive series manufacturing.

\subsection{Evaluation on products A-D (existing products)}
\begin{table}[hbtp]
	\centering
	\caption{Comparison of automated path planning results and designs from experts on existing products. The expert path is shown in the top row of Figure~\ref{fig:04_optvshuman}. The optimized paths are shown in the respective center and bottom row.}\label{tab:04_optvshuman}
	\begin{tabular}{K{30mm} | X{10mm} X{10mm} X{10mm} X{10mm} }
		\hline
		\rule{0pt}{11pt} & \rotatebox{90}{Product A~} & \rotatebox{90}{Product B~} & \rotatebox{90}{Product C~} & \rotatebox{90}{Product D~} \\
		\hline
		\rule{0pt}{11pt}\noindent Coverage of cooling surface & 94.3\,\%		& 98.3\,\% 	& 97.2\,\% 	& 95.5\,\% \\
		\hline
		\rule{0pt}{11pt}\noindent TIM overflow ratio & & & & \\
		\rule{0pt}{11pt} ~~~Expert path 		& 20.3\,\% 		& 13.1\,\%  & 19.5\,\% 	& 8.7\,\% \\
		\rule{0pt}{11pt} ~~~Optimized path 1 	& \textbf{7.1\,\%}		& \textbf{9.6\,\%} 	& \textbf{2.5\,\%}	& \textbf{4.3\,\%} \\
		\rule{0pt}{11pt} ~~~Optimized path 2	 & \textbf{7.1\,\%}  	& 12.1\,\% 	& 8\,\% 	& 5.9\,\% \\
		\hline
		\rule{0pt}{11pt}\noindent Number of segments & & & & \\
		\rule{0pt}{11pt} ~~~Expert path & 2 & 1 & 3 & 4 \\
		\rule{0pt}{11pt} ~~~Optimized path 1 & \textbf{1} & 1 & \textbf{1} & \textbf{1} \\
		\rule{0pt}{11pt} ~~~Optimized path 2 & \textbf{1} & 1 & \textbf{1} & \textbf{1} \\
	\end{tabular}
\end{table}

The paths which are gained from the automated path planning have further advantages.
The optimized paths all have a fixed material feedrate and they are not broken up into individual path segments.
This is an advantage in terms of cycle-time, since the dispense head does not need to start and stop the dispensing.
Having one continuous dispense path is furthermore beneficial with respect to process stability and quality inspection.

\subsection{Evaluation on product E (new product)}
\begin{figure}[hbt]
	\centering
	\includegraphics[width=0.7\columnwidth]{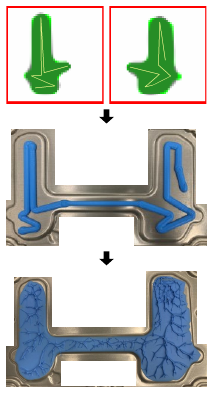}
	\caption{Optimization results, dispensed path and resulting coverage for the cooling surface for product E. In this case, our proposed approach has been used during the dispense pattern design (as opposed to product A-D, where the approach was not available yet). Only small adaptions to the optimized dispense path were made by the process expert. This evaluation was carried out under real-world conditions during a regular series ramp-up. This is a contrast to both Baeuerle et al.~\cite{baeuerle_rapid_2024} and Flaig et al.~\cite{flaig_practical_2023}, which provide results in a laboratory setting.}\label{fig:04_new_product}
\end{figure}

The underlying flow behavior model itself was validated on experimental data in a laboratory setting by Baeuerle et al.~\cite{baeuerle_rapid_2024}
The validation showed a good match with an error of the predicted coverage after joining being around 10\,\%.
Within this work, we validate an optimized dispense path on an actual product (product E).
The optimization result was given to the process experts, who made only minor adjustments.
The optimization output, the adjusted dispense pattern and the resulting coverage are shown in Figure~\ref{fig:04_new_product}.
For the optimization output at the top, the dispense path is shown in a light yellow color and the resulting coverage in a grey color.
It exhibits minor areas of the cooling surface not being covered.
This is due to large mechanical tolerances and partially no components being mounted in the areas below the heatsink.
The coverage has been assessed by the responsible engineers and found to meet the requirements well.
The results from the optimizer served as a very good starting point for the dispense path planning and greatly simplified the design process.

%

\subsection{Configuration of objective function}
A full report of the tested hyperparameter configurations is shown in the appendix.
The void area being part of the loss function showed a significant advantage over modeling the void presence in a binary fashion:
utilizing $w_{voidArea}$ instead of $w_{voidBin}$ consistently improved the convergence behavior by a large margin. 
Assigning a proportionally higher weight to $w_{init,over}$ as compared to $w_{comp, tab}$, $w_{voidBin}$ and $w_{voidArea}$ improves the convergence behavior. 
Assigning a proportionally lower weight to $w_{init,over}$ not only worsens the convergence behavior, but also slightly decreases the achieved cooling surface coverage. 
We did not see a clear advantage for any of the weighting strategies for the coverage area evaluation. 
Overall, the combination of setting $w_{voidArea}$ along with $w_{init,over}$ yielded best results with respect to both convergence properties and cooling surface coverage.

\subsection{Compliance with mechanical tolerances}\label{sec:compliancemechanicaltolerances}
\begin{figure}[hbt]
	\centering
	\includegraphics[width=0.9\columnwidth]{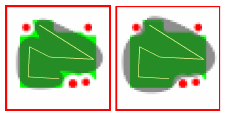}
	\caption{The proposed optimization approach yields a solution for product D, which complies with mechanical tolerances as follows: the resulting dispense path simultaneously matches the desired coverage conditions in a trade-off between both the maximum gap height and the minimum gap height. The left image shows the maximum gap setting, where a high coverage of the cooling surface (green) is achieved. The right image shows the minimum gap setting, where taboo zones (red) are being avoided. The shown dispense path - as indicated by the light yellow line - fulfills both criteria at the same time.}\label{fig:multigap}
\end{figure}

Figure~\ref{fig:multigap} shows the optimization result when including the mechanical tolerances during the formulation of the objective function.
The optimizer manages to find a dispense path, which fulfills the requirements well both for a maximum and minimum gap height after joining:
It manages to cover a large part of the cooling surface with the maximum gap height setting and to avoid taboo zones with the minimum gap height setting.

\section{Discussion}
The key to a successful optimization was to optimize the parameterized path as opposed to optimizing the initial material distribution (after dispensing/before joining).
This restricts the parameter space to feasible states with regard to the manufacturing constraints.
It significantly reduces the dimensionality of the parameter space, since instead of the full 2D distribution only a few path parameters are taken into account:
Even with a rather coarse discretization of $50\,\times\,50$ grid cells this would lead to 2\,500 degrees of freedom.
In contrast, a polygonal chain with 10 segments has only 20 degrees of freedom – assuming constant material feedrate.
Furthermore, additional constraints can be formulated straightforward.
This includes further constraints which have not been taken into account during this work yet, e.g. maximum or minimum segment length, minimum distance between individual segments or a rounding at the corners to reduce mechanical accelerations of the dispense head.

We see the choice of the parameter space to be optimized to be an issue with the studies of Kaufmann and Flaig et al. (\cite{kaufmann_optimized_2024, flaig_how_2023}).
They closely orient their optimization approach along the lines of classic topology optimization.
They optimize the material distribution after dispensing with the help of the flow behavior model of Kaufmann et al.~\cite{kaufmann_how_2023}
This effectively leads to the material being drawn back to the geometric center lines of the target surface and closely resembles the results of applying a medial axis transform.

We studied our optimization approach on TIM dispensing, but it is directly applicable to adhesive application.
This would benefit an even wider application range.

A constant feedrate as we implemented it during this work has advantages with regard to quality inspection.
An uninterrupted path with a high feedrate is further preferable in terms of cycle-time.
Creating a model for the cycle-time and integrating it into the objective function would be an extension of our approach, which would cover both objectives even better.
Furthermore, a model for the joining forces could be included to reduce material stress.

While we worked with a constant feedrate along the entire dispense path, a piece-wise constant or even varying feedrate can be realized easily.
This is a little more elaborate to implement on a manufacturing machine.
However, it creates additional degrees of freedom for the optimizer while still being within a feasible parameter range of the manufacturing equipment.
This leads to an even better area coverage.

We saw the formulation of boundary conditions as an objective rather than a hard constraint to be advantageous in terms of convergence properties:
reducing the void area rather than strictly preventing voids in a binary fashion showed clearly superior results.
This validates the findings of Ellefsen et al.~\cite{ellefsen_multiobjective_2017}, who proposed to formulate an objective with continuous values rather than an objective with hard constraints which strictly prohibit any violation.

Overall, the formulation of the objective function significantly affected the quality of the results and the convergence properties.
While we analyzed a sensibly chosen set of hyperparameters, we do not claim to have found the best hyperparameter setting yet.
Introducing an automated hyperparameter optimization could further benefit the setup.
However, since the hyperparameters are set for the optimization itself, the computation time for such an optimization would be quite high.

A limitation of our approach is the computation speed.
Adjusting the hyperparameters has improved the speed significantly.
However, we are running the optimizer many times for a single given product surface and it takes about one week to find a very well-tuned dispense path.
We run the optimization trials in parallel on each core of an INTEL Xeon E5-2680 processor.
Here, introducing the ANN as flow behavior model as proposed by Baeuerle et al.~\cite{baeuerle_rapid_2024} could help.
This would not only decrease the computation time of the gradient-free optimization, but would also enable the use of gradient-based optimizers.

Our comparison of the expert path with the optimized path is carried out using the heuristic flow behavior model.
An experimental evaluation on product level would of course make this part of our evaluation much stronger.
However, the efforts for such experiments are quite high.
It requires not only full product replicas, but also additional equipment to carry out the joining process in a setup, that is equivalent to the conditions of series manufacturing.
While we were not able to carry out this type of validation, we emphasize once again the validations we have done: A) The flow behavior model of Baeuerle et al.~\cite{baeuerle_rapid_2024} was validated in a laboratory setting.
B) We have introduced an optimized pattern during the series ramp up of a new product.
The shown results are obtained from actual series manufacturing equipment.

\section{Conclusion}
We extend the research field of Coverage Path Planning beyond constant coverage path widths towards a complex modeling of the area coverage.
We propose a new optimization approach specifically for the dispensing of Thermal Interface Material.
It is transferable to other dispensing applications such as adhesive dispensing.
We lay out the needed configuration in detail.
We validate our approach against existing manual dispense path planning and present experimental results not only in a laboratory setting but from actual series manufacturing equipment.
Our approach fully complies with the essential manufacturing constraints and can easily be extended to further constraints.
We plan to extend our approach towards other flow behavior models – specifically towards the ANN proposed by Baeuerle et al.~\cite{baeuerle_rapid_2024}
This includes especially the study of second-order optimization methods and also the formulation of further constraints with respect to the dispense path.

\appendix
\section{}
Table~\ref{tab:apx_optresults} shows the results of our hyperparameter study. It includes various weighting strategies. The strategies are evaluated with the a) the convergence ratio, b) the coverage percentage of the cooling surface and c) the average of both as reported in the last columns. The following parameters remain constant: $w_{comp, cool}\,=\,0$, $w_{comp, over}\,=\,1$.

\begin{table*}[hbtp]
	\centering
	\caption{Parameter study for the components of the objective function. Good result values are indicated with a stronger blue color. Relevant violations w.r.t. voids and taboo zones are considered in the convergence ratio. The column \textit{average performance} is the average of the previous two columns. This analysis has been carried out using product D.}\label{tab:apx_optresults}
	\begin{tabular}{E{12mm}  E{12mm} E{12mm} E{12mm} E{12mm} E{12mm} E{12mm} E{12mm} E{12mm} E{12mm} E{12mm} E{12mm} E{12mm}}
\rule{0pt}{11pt} \rotatebox{90}{\parbox{1.8cm}{{w\_comp,tab}}} &  \rotatebox{90}{\parbox{1.8cm}{{w\_voidArea}}} &  \rotatebox{90}{\parbox{1.8cm}{{w\_voidBin}}} &  \rotatebox{90}{\parbox{1.8cm}{{w\_init,over}}} &  \rotatebox{90}{\parbox{1.8cm}{{f\_con}}} &  \rotatebox{90}{\parbox{1.8cm}{{f\_area}}} &  \rotatebox{90}{\parbox{1.8cm}{{f\_init}}} &  \rotatebox{90}{\parbox{1.8cm}{{Coverage ratio of cooling surface [\%]}}} &  \rotatebox{90}{\parbox{1.8cm}{{Convergence ratio [\%]}}} &  \rotatebox{90}{\parbox{1.8cm}{{Average performance [\%]}}}\\ 
\hline 
\rule{0pt}{8pt}    100 &    100 &      0 &      0 &    con &    con &   none & \cellcolor{blue!4}  0.89 & \cellcolor{blue!28}  0.82 & \cellcolor{blue!26}  0.85\\ 
\rule{0pt}{6pt}   1000 &   1000 &      0 &      0 &    con &    con &   none & \cellcolor{blue!18}  0.91 & \cellcolor{blue!25}  0.78 & \cellcolor{blue!24}  0.85\\ 
\rule{0pt}{6pt}  10000 &  10000 &      0 &      0 &    con &    con &   none & \cellcolor{blue!16}  0.91 & \cellcolor{blue!19}  0.71 & \cellcolor{blue!19}  0.81\\ 
\rule{0pt}{6pt}    100 &    100 &      0 &      0 &    log &    con &   none & \cellcolor{blue!26}  0.92 & \cellcolor{blue!25}  0.78 & \cellcolor{blue!25}  0.85\\ 
\rule{0pt}{6pt}   1000 &   1000 &      0 &      0 &    log &    con &   none & \cellcolor{blue!21}  0.91 & \cellcolor{blue!21}  0.74 & \cellcolor{blue!22}  0.83\\ 
\rule{0pt}{6pt}  10000 &  10000 &      0 &      0 &    log &    con &   none & \cellcolor{blue!21}  0.91 & \cellcolor{blue!26}  0.79 & \cellcolor{blue!26}  0.85\\ 
\rule{0pt}{6pt}    100 &    100 &      0 &    100 &    log &    con &    lin & \cellcolor{blue!32}  0.93 & \cellcolor{blue!28}  0.82 & \cellcolor{blue!29}  0.87\\ 
\rule{0pt}{6pt}    100 &    100 &      0 &   1000 &    log &    con &    lin & \cellcolor{blue!35}  0.93 & \cellcolor{blue!30}  0.84 & \cellcolor{blue!31}  0.88\\ 
\rule{0pt}{6pt}    100 &    100 &      0 &    100 &    log &    con &    log & \cellcolor{blue!34}  0.93 & \cellcolor{blue!27}  0.81 & \cellcolor{blue!28}  0.87\\ 
\rule{0pt}{6pt}    100 &    100 &      0 &   1000 &    log &    con &    log & \cellcolor{blue!36}  0.93 & \cellcolor{blue!42}  0.95 & \cellcolor{blue!42}  0.94\\ 
\rule{0pt}{6pt}    100 &      0 &    100 &      0 &    log &    con &   none & \cellcolor{blue!2}  0.88 & \cellcolor{blue!0}  0.26 & \cellcolor{blue!0}  0.57\\ 
\rule{0pt}{6pt}   1000 &      0 &   1000 &      0 &    log &    con &   none & \cellcolor{blue!14}  0.90 & \cellcolor{blue!0}  0.27 & \cellcolor{blue!0}  0.59\\ 
\rule{0pt}{6pt}    100 &    100 &      0 &      0 &    log &    lin &   none & \cellcolor{blue!25}  0.92 & \cellcolor{blue!24}  0.77 & \cellcolor{blue!24}  0.84\\ 
\rule{0pt}{6pt}    100 &    100 &      0 &      0 &    log &    squ &   none & \cellcolor{blue!5}  0.89 & \cellcolor{blue!22}  0.75 & \cellcolor{blue!20}  0.82\\ 
\rule{0pt}{6pt}    100 &    100 &      0 &      0 &    log &    log &   none & \cellcolor{blue!7}  0.89 & \cellcolor{blue!21}  0.74 & \cellcolor{blue!20}  0.82\\ 
\rule{0pt}{6pt}    100 &    100 &      0 &      0 &    log &    con &   none & \cellcolor{blue!31}  0.93 & \cellcolor{blue!25}  0.78 & \cellcolor{blue!26}  0.85\\ 
\rule{0pt}{6pt}    100 &    100 &      0 &    100 &    con &    con &    lin & \cellcolor{blue!26}  0.92 & \cellcolor{blue!34}  0.88 & \cellcolor{blue!34}  0.90\\ 
\rule{0pt}{6pt}    100 &    100 &      0 &   1000 &    con &    con &    lin & \cellcolor{blue!36}  0.93 & \cellcolor{blue!41}  0.94 & \cellcolor{blue!41}  0.93\\ 
\rule{0pt}{6pt}    100 &    100 &      0 &    100 &    con &    con &    log & \cellcolor{blue!29}  0.92 & \cellcolor{blue!28}  0.82 & \cellcolor{blue!29}  0.87\\ 
\rule{0pt}{6pt}    100 &    100 &      0 &   1000 &    con &    con &    log & \cellcolor{blue!31}  0.92 & \cellcolor{blue!34}  0.88 & \cellcolor{blue!34}  0.90\\ 
\rule{0pt}{6pt}    100 &    100 &      0 &      0 &    con &    con &   none & \cellcolor{blue!31}  0.93 & \cellcolor{blue!29}  0.83 & \cellcolor{blue!30}  0.88\\ 
\rule{0pt}{6pt}    100 &    100 &      0 &      0 &    con &    lin &   none & \cellcolor{blue!13}  0.90 & \cellcolor{blue!21}  0.74 & \cellcolor{blue!21}  0.82\\ 
\rule{0pt}{6pt}    100 &    100 &      0 &      0 &    con &    squ &   none & \cellcolor{blue!8}  0.89 & \cellcolor{blue!28}  0.82 & \cellcolor{blue!26}  0.86\\ 
\rule{0pt}{6pt}    100 &    100 &      0 &      0 &    con &    log &   none & \cellcolor{blue!6}  0.89 & \cellcolor{blue!22}  0.75 & \cellcolor{blue!21}  0.82\\ 
\rule{0pt}{6pt}   1000 &   1000 &      0 &   1000 &    log &    con &    lin & \cellcolor{blue!27}  0.92 & \cellcolor{blue!32}  0.86 & \cellcolor{blue!32}  0.89\\ 
\rule{0pt}{6pt}   1000 &   1000 &      0 &  10000 &    log &    con &    lin & \cellcolor{blue!32}  0.93 & \cellcolor{blue!32}  0.86 & \cellcolor{blue!33}  0.89\\ 
\rule{0pt}{6pt}   1000 &   1000 &      0 &   1000 &    log &    con &    log & \cellcolor{blue!27}  0.92 & \cellcolor{blue!30}  0.84 & \cellcolor{blue!30}  0.88\\ 
\rule{0pt}{6pt}   1000 &   1000 &      0 &  10000 &    log &    con &    log & \cellcolor{blue!32}  0.93 & \cellcolor{blue!36}  0.90 & \cellcolor{blue!36}  0.91\\ 
\rule{0pt}{6pt}    100 &      0 &    100 &      0 &    con &    con &   none & \cellcolor{blue!2}  0.88 & \cellcolor{blue!0}  0.22 & \cellcolor{blue!0}  0.55\\ 
\rule{0pt}{6pt}   1000 &      0 &   1000 &      0 &    con &    con &   none & \cellcolor{blue!4}  0.89 & \cellcolor{blue!0}  0.21 & \cellcolor{blue!0}  0.55\\ 
\rule{0pt}{6pt}   1000 &      0 &   1000 &      0 &    log &    con &   none & \cellcolor{blue!0}  0.86 & \cellcolor{blue!0}  0.20 & \cellcolor{blue!0}  0.53\\ 
\rule{0pt}{6pt}  10000 &      0 &  10000 &      0 &    log &    con &   none & \cellcolor{blue!3}  0.88 & \cellcolor{blue!0}  0.25 & \cellcolor{blue!0}  0.57\\ 
\rule{0pt}{6pt}   1000 &   1000 &      0 &      0 &    con &    con &   none & \cellcolor{blue!10}  0.90 & \cellcolor{blue!23}  0.76 & \cellcolor{blue!22}  0.83\\ 
\rule{0pt}{6pt}   1000 &   1000 &      0 &      0 &    con &    lin &   none & \cellcolor{blue!23}  0.92 & \cellcolor{blue!26}  0.79 & \cellcolor{blue!26}  0.85\\ 
\rule{0pt}{6pt}   1000 &   1000 &      0 &      0 &    con &    squ &   none & \cellcolor{blue!6}  0.89 & \cellcolor{blue!17}  0.68 & \cellcolor{blue!16}  0.78\\ 
\rule{0pt}{6pt}   1000 &   1000 &      0 &      0 &    con &    log &   none & \cellcolor{blue!4}  0.89 & \cellcolor{blue!19}  0.71 & \cellcolor{blue!18}  0.80\\ 
\rule{0pt}{6pt}   1000 &   1000 &      0 &      0 &    log &    con &   none & \cellcolor{blue!9}  0.90 & \cellcolor{blue!21}  0.73 & \cellcolor{blue!20}  0.81\\ 
\rule{0pt}{6pt}   1000 &   1000 &      0 &      0 &    log &    lin &   none & \cellcolor{blue!10}  0.90 & \cellcolor{blue!25}  0.78 & \cellcolor{blue!23}  0.84\\ 
\rule{0pt}{6pt}   1000 &   1000 &      0 &      0 &    log &    squ &   none & \cellcolor{blue!0}  0.87 & \cellcolor{blue!14}  0.64 & \cellcolor{blue!12}  0.76\\ 
\rule{0pt}{6pt}   1000 &   1000 &      0 &      0 &    log &    log &   none & \cellcolor{blue!14}  0.91 & \cellcolor{blue!25}  0.78 & \cellcolor{blue!24}  0.84\\ 
\rule{0pt}{6pt}     10 &     10 &      0 &     10 &    con &    con &    lin & \cellcolor{blue!40}  0.93 & \cellcolor{blue!25}  0.78 & \cellcolor{blue!26}  0.86\\ 
\rule{0pt}{6pt}     10 &     10 &      0 &     10 &    con &    lin &    lin & \cellcolor{blue!22}  0.92 & \cellcolor{blue!25}  0.78 & \cellcolor{blue!25}  0.85\\ 
\rule{0pt}{6pt}     10 &     10 &      0 &     10 &    con &    squ &    lin & \cellcolor{blue!24}  0.92 & \cellcolor{blue!21}  0.73 & \cellcolor{blue!21}  0.82\\ 
\rule{0pt}{6pt}     10 &     10 &      0 &     10 &    con &    log &    lin & \cellcolor{blue!26}  0.92 & \cellcolor{blue!27}  0.81 & \cellcolor{blue!28}  0.86\\ 
\rule{0pt}{6pt}     10 &     10 &      0 &     10 &    log &    con &    lin & \cellcolor{blue!42}  0.93 & \cellcolor{blue!31}  0.85 & \cellcolor{blue!33}  0.89\\ 
\rule{0pt}{6pt}     10 &     10 &      0 &     10 &    log &    lin &    lin & \cellcolor{blue!26}  0.92 & \cellcolor{blue!31}  0.85 & \cellcolor{blue!31}  0.88\\ 
\rule{0pt}{6pt}     10 &     10 &      0 &     10 &    log &    squ &    lin & \cellcolor{blue!23}  0.92 & \cellcolor{blue!28}  0.82 & \cellcolor{blue!28}  0.87\\ 
\rule{0pt}{6pt}     10 &     10 &      0 &     10 &    log &    log &    lin & \cellcolor{blue!28}  0.92 & \cellcolor{blue!32}  0.86 & \cellcolor{blue!32}  0.89\\ 
\rule{0pt}{6pt}    100 &    100 &      0 &     10 &    con &    con &    lin & \cellcolor{blue!23}  0.92 & \cellcolor{blue!26}  0.79 & \cellcolor{blue!26}  0.85\\ 
\rule{0pt}{6pt}    100 &    100 &      0 &     10 &    con &    lin &    lin & \cellcolor{blue!9}  0.90 & \cellcolor{blue!21}  0.74 & \cellcolor{blue!20}  0.82\\ 
\rule{0pt}{6pt}    100 &    100 &      0 &     10 &    con &    squ &    lin & \cellcolor{blue!11}  0.90 & \cellcolor{blue!21}  0.74 & \cellcolor{blue!21}  0.82\\ 
\rule{0pt}{6pt}    100 &    100 &      0 &     10 &    con &    log &    lin & \cellcolor{blue!12}  0.90 & \cellcolor{blue!28}  0.82 & \cellcolor{blue!27}  0.86\\ 
\rule{0pt}{6pt}    100 &    100 &      0 &     10 &    log &    con &    lin & \cellcolor{blue!14}  0.90 & \cellcolor{blue!25}  0.78 & \cellcolor{blue!24}  0.84\\ 
\rule{0pt}{6pt}    100 &    100 &      0 &     10 &    log &    lin &    lin & \cellcolor{blue!9}  0.90 & \cellcolor{blue!24}  0.77 & \cellcolor{blue!23}  0.83\\ 
\rule{0pt}{6pt}    100 &    100 &      0 &     10 &    log &    squ &    lin & \cellcolor{blue!6}  0.89 & \cellcolor{blue!19}  0.71 & \cellcolor{blue!18}  0.80\\ 
\rule{0pt}{6pt}    100 &    100 &      0 &     10 &    log &    log &    lin & \cellcolor{blue!9}  0.90 & \cellcolor{blue!21}  0.74 & \cellcolor{blue!20}  0.82\\ 
\end{tabular}
\end{table*}
%
%
\section*{Acknowledgment}
We thank Ralph Nyilas and Barnabas Szilagyi for the valuable exchange with respect to the dispensing process and the properties of the manufacturing equipment.
\section*{Author statement}
We describe the individual contributions of Simon Baeuerle (SB), Andreas Steimer (AS) and Ralf Mikut (RM) using CRediT~\cite{brand_beyond_2015}: \textit{Writing - Original Draft}: SB; \textit{Writing - Review \& Editing}: AS, RM; \textit{Conceptualization}: SB, AS, RM; \textit{Investigation}: SB; \textit{Methodology}: SB, AS, RM; \textit{Software}: SB; \textit{Supervision}: AS, RM; \textit{Project Administration}: RM; \textit{Funding Acquisition}: SB, RM.

\clearpage
%
%
%
%
%
\bibliographystyle{IEEEtran}
\bibliography{literature}
\ifCLASSOPTIONcaptionsoff
  \newpage
\fi

%

%

%




\end{document}